\documentclass[12pt]{article}

\newcommand{\no}{\nonumber}
\newcommand{\eps}{\varepsilon}

\usepackage{graphicx}
\usepackage{dcolumn}
\usepackage{bm}

\begin{document}

\large

\begin{center}

{\Large\bf
Ultra-large-scale electronic structure theory \\
 and numerical algorithm
}\\[5mm]

{
\underline{T. Hoshi}$^{1,2}$
 }\\

\vspace{2mm}

{\normalsize\em
$^1$Department of Applied Mathematics and Physics, Tottori University; \\
$^2$Core Research for Evolutional Science and Technology, 
Japan Science and Technology Agency (CREST-JST)
 }
\end{center}

This article is composed of two parts;
In the first part (Sec. 1), 
the ultra-large-scale electronic structure theory 
is reviewed
for (i) its fundamental numerical algorithm and 
(ii) its role in nano-material science. 
The second part (Sec. 2) is devoted 
to the mathematical foundation of 
the large-scale electronic structure theory
and their numerical aspects.

\section{Large-scale electronic structure theory and nano-material science}

\subsection{Overview}

Nowadays
electronic structure theory
gives a microscopic foundation of material science 
and  provides 
atomistic simulations  in which
electrons are treated as wavefunction within quantum mechanics.
An example is given in the upper left panel of Fig.1.
For years, 
we have developed fundamental theory and program code 
for large-scale electronic structure calculations, particularly,
for nano materials.  [1-6]
The code was applied to several nano materials
with $10^2$-$10^7$ atoms,
whereas 
standard electronic structure calculations 
are carried out typically with $10^2$ atoms.
Two application studies, for silicon and gold, are shown in 
the right panels and the lower left panels of Fig.1, respectively.   
Now the code is being reorganized as a simulation package, 
named as ELSES (Extra-Large-Scale Electronic Structure calculation), 
for a wider range of users and applications in science and industry. [6]

\begin{figure}[hb]
\begin{center}
\includegraphics[width=13cm]{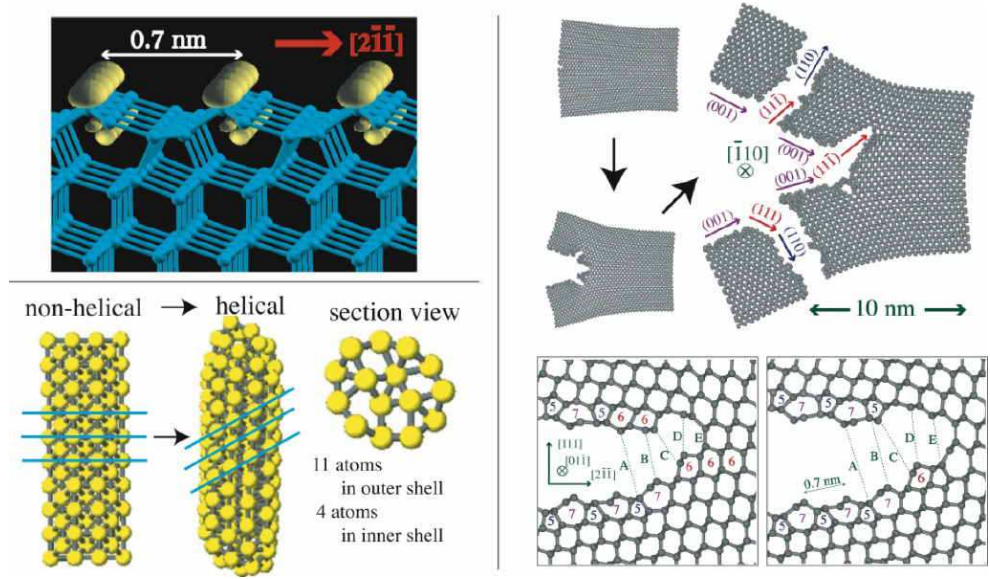}
\end{center}
\caption{
Upper left panel: Example of calculated electronic wavefunction on a silicon surface 
 (A \lq $\pi$-type' electron state on Si(111)-2x1 surface, 
 given by a standard electronic structure calculation). 
Right panels: Application of our code
to fracture dynamics of silicon crystal. [2]
In results, the fracture path is bent into 
experimentally-observed planes (right upper panels) and 
and reconstructed Si(111)-2x1 surfaces appear with step formation
(right lower panels).
Lower left panels: 
Application of our code to 
formation process of helical multishell gold nanowire [5]
that was reported experimentally. [7] 
A non-helical structure is transformed into a helical one
(left and middle panels). 
The section view (right panel) shows 
a  multishell structure, called  \lq 11-4 structure', 
in which the outer and inner shells consist of eleven and four atoms, 
respectively.
 }
\label{PHOTO}
\end{figure}%


\subsection{Methodology}

Our methodologies contain several mathematical theories, 
as Krylov-subspace theories for large sparse matrices. 
Here we focus on a solver method of shifted linear equations,
called \lq shifted conjugate-orthogonal 
conjugate-gradient (COCG) method' [3][4]. 
A quantum mechanical calculation within our studies 
is conventionally reduced  to an eigen-value problem with 
a real-symmetric $N \times N$ matrix $H$ 
(Hamiltonian matrix),
which will cost an $O(N^3)$ computational time. 
In our method, instead, 
the problem is reduced to a set of shifted linear equations;
\begin{eqnarray}
( z^{(k)} I - H ) \, \, \bm{x}(z^{(k)}) = \bm{b}
 \label{EQ1}
\end{eqnarray}
with a set of given complex variables $\{ z^{(1)}, z^{(2)},....z^{(L)} \}$ 
that have physical meaning of energy points. 
See Sec.~2 for mathematical foundation.  
Since the matrix $( z^{(k)} I - H )$ is complex symmetric,
one can solve the equations 
by the COCG method, 
independently among the energy points. [8]
In these calculations, 
the procedure of matrix-vector multiplications governs
the computational time. 

For the problem of Eq.(1), a novel Krylov-subspace algorithm,
the shifted COCG method,  was constructed [3][4],  
in which we  should solve the equation actually 
only at one energy point (reference system). 
The solutions of the other energy points (shifted systems) 
can be given without any matrix-vector multiplication,
which leads to a drastic reduction of computational time. 
The key feature of the shifted COCG method stems from 
the fact that the residual vectors 
$\bm{r}^{(k)} \equiv ( z^{(k)} I - H ) \bm{x}(z^{(k)}) - \bm{b}$
are collinear among energy points,
owing to the theorem of collinear residuals. [9]

Moreover, the shifted COCG method gives another drastic reduction 
of computational time,
when one does not need all the elements of the solution vector $\bm{x}(z^{(k)})$,
as in many cases of our studies.  [3]
For example, the inner product 
\begin{eqnarray}
\rho(\bm{b}, z^{(k)})\equiv (\bm{b}, \bm{x}(z^{(k)}))
\label{EQ2}
\end{eqnarray}
is particularly interested in our cases.
The shifted COCG method gives an iterative algorithm 
for the scaler $(\bm{b}, \bm{x}(z^{(k)}))$,
without calculating the vector $\bm{x}(z^{(k)})$,  
among the shifted systems. 
The quantity of Eq.(2) is known as \lq local density of states' (See Sec.~2). 
It is an energy-resolved electron distribution at a point in real space 
and can be measured experimentally
as a bias-dependent image 
of scanning tunneling microscope
(See textbooks of condensed matter physics). 


\section{Note on mathematical formulation}

Here a brief note is devoted to mathematical relationship
between eigen-value equation and shifted linear equation 
in the electronic structure theory. 
A notation, known as bra-ket notation in quantum mechanics, is used. 
See Appendix for the details of the notation.

\subsection{Original problem}

Our problem in
electronic structure calculation is, conventionally, reduced to 
an eigen-value problem, an effective Schr\"odinger equation, 
\begin{eqnarray}
H | v_\alpha \rangle = \eps_\alpha  | v_\alpha \rangle,
\label{EQ-SHRODINGER}
\end{eqnarray}
where $H$ is a given $N \times N$ real-symmetric matrix,
called Hamiltonian matrix. 
Eigen vectors form a complete orthogonal basis set;
\begin{eqnarray}
& & \langle v_\alpha | v_\alpha \rangle = \delta_{\alpha \beta}
\label{EQ-ORTHO} \\
& &  \sum_\alpha | v_\alpha \rangle  \langle v_\alpha | = I
\label{EQ-COMP} 
\end{eqnarray}
where $I$ is the unit matrix. 

In actual calculation, the matrix $H$ is sparse. 
Each basis of the matrix and the vectors
corresponds to the wavefunction localized in real space.
Hereafter physical discussion is given in the case that 
the physical system has $N$ atoms and 
only one basis is considered for one atom.
In short, 
the $i$-th basis ($i=1,2,3...N$) 
corresponds to the basis localized on the $i$-th atom.
Moreover we suppose, for simplicity,  
that the eigen values are not degenerated
($\eps_1 <  \eps_2 < \eps_3 ......$). 

On the other hand,
the linear equation of Eq.~(\ref{EQ1}) is rewritten in the present notation; 
\begin{eqnarray}
(z - H) | x_j(z) \rangle  = | j \rangle 
\label{EQ-SHIFTED-LE}
\end{eqnarray}
with a complex valuable $z \equiv \eps  + i \eta$.
The valuable $z$ corresponds to the energy
with a tiny imaginary part $\eta$ ($\eta \rightarrow +0$).

The purpose within the present  calculation procedure
is to obtain selected elements of the following matrix $D$; 
\begin{eqnarray}
D(\eps) \equiv  \delta(\eps - H) =  
\sum_\alpha | v_\alpha \rangle \delta(\eps - \eps_\alpha) \langle v_\alpha |
\label{EQ-HAT-DESP-ORG}
\end{eqnarray}
or
\begin{eqnarray}
D_{ij}(\eps) =  
\sum_\alpha \langle i  | v_\alpha \rangle \delta(\eps - \eps_\alpha)
 \langle v_\alpha | j \rangle. 
\label{EQ-HAT-DESP-ORG2}
\end{eqnarray}
This matrix is called 
density-of-states (DOS) matrix,
since its trace  
\begin{eqnarray}
{\rm Tr}[D(\eps) ]
= \sum_\alpha \delta(\eps - \eps_\alpha),
\label{EQ-DESP}
\end{eqnarray}
is called density of states. 
The physical meaning of Eq.~(\ref{EQ-DESP}) 
is the spectrum of eigen-value distribution.

In actual numerical calculation, 
the delta function in Eqs.~(\ref{EQ-HAT-DESP-ORG}) and (\ref{EQ-HAT-DESP-ORG2})
is replaced by an analytic function, a \lq smoothed' delta function, 
since  numerical calculation should be 
free from the singularity of the exact delta function ($\delta(0)=\infty$). 
The \lq smoothed' delta function is defined as
\begin{eqnarray}
\delta_{\eta}(\eps)  \equiv
-\frac{1}{\pi} {\rm Im} \left[ \frac{1}{ \eps+ i \eta  } \right]  
= \frac{1}{\pi} \frac{\eta}{ \eps ^2 + \eta^2} 
\label{EQ-DELTA-FN}
\end{eqnarray}
with a finite positive value of $\eta (> 0)$.
The smoothed delta function gives the exact delta function 
in the limit of 
\begin{eqnarray}
\lim_{\eta \rightarrow +0} \delta_{\eta}(\eps)  
= \delta(\eps). 
\end{eqnarray}
The physical meaning of $\eta$ is 
the width of the \lq smoothed' delta function 
$\delta_{\eta}(\eps)$.
Hereafter, the DOS matrix is defined as
\begin{eqnarray}
D(\eps) \equiv  \delta_{\eta}(\eps - H) =  
\sum_\alpha | v_\alpha \rangle \delta_{\eta}(\eps - \eps_\alpha) \langle v_\alpha |
\label{EQ-HAT-DESP}
\end{eqnarray}
or
\begin{eqnarray}
D_{ij}(\eps) =  
\sum_\alpha \langle i  | v_\alpha \rangle \delta_{\eta}(\eps - \eps_\alpha)
 \langle v_\alpha | j \rangle. 
\label{EQ-HAT-DESP2}
\end{eqnarray}
It is noteworthy that 
the smoothed delta function has a finite maximum  
\begin{eqnarray}
\delta_{\eta}(0)= \frac{1}{\eta}
\end{eqnarray}
and its integration gives the unity
\begin{eqnarray}
\int_{-\infty}^{\infty} \delta_{\eta}(\eps) d \eps = 
\frac{1}{\pi} \left[ {\rm tan}^{-1} 
\left(\frac{\eps}{\eta} \right) \right]_{\eps=-\infty}^{\eps=\infty} = 1. 
\end{eqnarray}

\subsection{Green's function and calculation procedure}

The Green's function is defined as an inverse matrix of
\begin{eqnarray}
G(z) \equiv \frac{1}{z - H} =
\sum_\alpha \frac{ | v_\alpha \rangle \langle v_\alpha |}{ z  - \varepsilon_\alpha } . 
\end{eqnarray}
The solution of Eq.~(\ref{EQ-SHIFTED-LE})
gives the Green's function;
\begin{eqnarray}
G_{ij}(z) \equiv \langle i | G(z) | j \rangle  =   \langle i | x_j(z) \rangle
\label{EQ-X-G}
\end{eqnarray}
and the DOS matrix is given from the Green's function; 
\begin{eqnarray}
D(\eps) = -\frac{1}{\pi} {\rm Im} \left[ G(\eps + i \eta)\right], 
\label{EQ-G-D}
\end{eqnarray}
under the relations of Eq.~(\ref{EQ-DELTA-FN}).
In conclusion, 
the calculation procedure,
from the Hamiltonian matrix to the DOS matrix,  is illustrated, as follows;
\begin{eqnarray}
H 
\stackrel{(\ref{EQ-SHIFTED-LE})(\ref{EQ-X-G})}{\Longrightarrow} 
 G 
 \stackrel{(\ref{EQ-G-D})}{\Longrightarrow}
 D
\end{eqnarray}

\subsection{Physical quantities for energy decomposition}

Now we present two quantities, 
local density of states (LDOS) and crystal orbital Hamiltonian population (COHP)[10],
as examples of important physical quantities that is calculated
from selected element of the DOS matrix. 
These quantities appear
in the decomposition methods of
the electronic structure energy.
The electronic structure energy is defined as
\begin{eqnarray}
E = \sum_\alpha \eps_\alpha f(\eps_\alpha), 
\label{EQ-ESUM}
\end{eqnarray}
where $f(\eps_\alpha)$ is the number of electrons
that occupy the wavefunction with the energy of $\eps_\alpha$.
In the present case, the case of zero temperature,
the function is reduced to  a step-function form of 
\begin{eqnarray}
 f(\eps) \equiv  \theta(\mu - \eps)
\end{eqnarray}
with a given value of $\mu$ (chemical potential).
Eqs.(\ref{EQ-ESUM}), (\ref{EQ-DESP}) lead us to 
the expression of the energy with DOS; 
\begin{eqnarray}
E &=& \int_{-\infty}^{\infty}  f(\eps) \, \eps \,
 \sum_\alpha \delta(\eps - \eps_\alpha) \, d \eps \no \\
   &=& \int_{-\infty}^{\infty}  f(\eps) \, \eps \, {\rm Tr}[D(\eps)] \, d \eps
\label{EQ-ESUM2}
\end{eqnarray}

\subsubsection{Decomposition with local density of states}

LDOS is defined as diagonal elements 
\begin{eqnarray}
n_{i}(\eps) \equiv \langle i | D(\eps) | i  \rangle 
= 
\sum_\alpha | \langle i | v_\alpha \rangle| ^2  \delta(\eps - \eps_\alpha) 
\label{EQ-LDOS}
\end{eqnarray}
of the DOS matrix and 
the energy is decomposed into 
the contributions of LDOS, $\{ n_i(\eps) \}_i$;
\begin{eqnarray}
E  = \sum_i \int_{-\infty}^{\infty}  f(\eps)  \eps  n_i(\eps)  d \eps.
\end{eqnarray}
Physical meaning of LDOS is a weighted eigen-value distribution;
For example, 
if an eigen vector $| v_\alpha \rangle$ has a large weight 
on the $i$-th basis, 
the local DOS $n_i(\eps)$ has a large peak
at the energy level of $\eps=\eps_\alpha$. 
The LDOS $n_i(\eps)$ corresponds 
to the experimental image 
of the scanning tunneling microscope
(See the end of Sec. 1).

\subsubsection{Decomposition with crystal orbital Hamiltonian population}

Another decomposition of the energy can be derived with an expression of
\begin{eqnarray}
E  =\int_{-\infty}^{\infty}  f(\eps)  {\rm Tr}[D(\eps)H]  d \eps.
\label{EQ-ESUM-DH}
\end{eqnarray}
Eq.(\ref{EQ-ESUM-DH}) is proved from 
the first line of Eq.(\ref{EQ-ESUM2}) and the relation of
\begin{eqnarray}
{\rm Tr}[D(\eps)H]  &=& {\rm Tr}[\delta_\eta(\eps - H)H] \no \\
 &=& \sum_\alpha \langle \phi_\alpha |  \delta_\eta(\eps - H)H | \phi_\alpha \rangle \no \\
 &=& \sum_\alpha \langle \phi_\alpha |  \delta_\eta(\eps - \eps_\alpha)
                   \eps_\alpha | \phi_\alpha \rangle \no \\
 &=& \sum_\alpha \delta_\eta(\eps - \eps_\alpha) \eps_\alpha \no \\
 &=& \eps \sum_\alpha \delta_\eta(\eps - \eps_\alpha). 
\end{eqnarray}
The last equality is satisfied by the exact delta function ($\eta \rightarrow 0+$).
Eq.(\ref{EQ-ESUM-DH}) gives another decomposition of the energy
\begin{eqnarray}
E  = \sum_{ij} 
\int_{-\infty}^{\infty}  f(\eps)  C_{ij}(\eps)  d \eps, 
\label{EQ-ESUM-COHP}
\end{eqnarray}
where the matrix $C$ is defined as
\begin{eqnarray}
C_{ij}(\eps)  \equiv D_{ij}(\eps) H_{ji}
\label{COHP}
\end{eqnarray}
and is called crystalline orbital Hamiltonian population (COHP) [10].
The physical meaning of COHP 
is an energy spectrum of electronic wavefunctions, 
or \lq chemical bond', 
that lie between $i$-th and $j$-th bases. 
See the papers [3, 10] for details.

\subsection{Numerical aspects with Krylov subspace theory}

Several numerical aspects are discussed 
for the calculated quantities within Krylov subspace theory,
such as the shifted COCG algorithm. 
LDOS is focused on, as an example. 
When we solve the shifted linear equation of
Eq.~(\ref{EQ-SHIFTED-LE})
within the $\nu$-th order Krylov subspace
\begin{eqnarray}
K^{(\nu)}(H; | j \rangle ) \equiv {\rm span} 
\left\{ | j \rangle , H | j \rangle , H^2| j \rangle , ......,H^{\nu-1} | j \rangle \right\},
\end{eqnarray}
the resultant LDOS should include deviation
from the properties in the previous sections, 
properties of the exact solution. 

First, 
the number of peaks in the LDOS function ($n_i(\eps)$)
is equal to $\nu$, the dimension of Krylov subspace,  
whereas the number of peaks in 
the exact solution, given in Eq.(\ref{EQ-LDOS}) 
is $N$, the dimension of the original matrix.  
A typical behavior is seen Fig.3(a) of Ref.[3],
a case with $\nu=30$ and $N=1024$. 
Here we note that 
the calculation with such a small subspace
($\nu=30$) gives  satisfactory results
in several physical quantities [3],
mainly because many physical quantities are defined
by a contour integral with respect to the energy, 
as in Eq.~(\ref{EQ-ESUM2}),  
and the information of individual peaks is not essential.

Second,
the calculated function $n_i(\eps)$ in the Krylov subspace can be negative
$(n_i(\eps) < 0)$, 
whereas the exact one, given in Eq.(\ref{EQ-LDOS}), 
is always positive. 
In the exact solution of Eq.(\ref{EQ-LDOS}), 
peaks in $n_i(\eps)$ are given by
the poles of the Green's function $G(z)$ 
($z = \eps_1, \eps_2,....$)
and the function $n_i(\eps)$ contains smoothed delta functions of
\begin{eqnarray}
\delta_{\eta}(\eps - \eps_\alpha)  \equiv
-\frac{1}{\pi} {\rm Im} \left[ \frac{1}{ (\eps - \eps_\alpha) + i \eta  } \right].  
\end{eqnarray}
Since $\eps_\alpha$ is an eigen value of the real-symmetric matrix $H$
and is real $({\rm Im}[\eps_\alpha] =0)$, 
the exact Green's function $G(z)$ has 
poles only on real axis. 
In the calculation within Krylov subspace, however, 
the poles can be deviated from real axis
($\eps_\alpha = \eps_\alpha^{\rm (r)}+ i \eps_\alpha^{\rm (i)}, 
\eps_\alpha^{\rm (i)} \ne 0$).
In that case, the sign of the smoothed delta function can be negative;
\begin{eqnarray}
\delta_{\eta}(\eps - \eps_\alpha)  &\equiv&
-\frac{1}{\pi} {\rm Im} \left[ \frac{1}{ (\eps - \eps_\alpha) + i \eta  } \right]  \no \\
&=&
-\frac{1}{\pi} {\rm Im} \left[ \frac{1}{ (\eps - \eps_\alpha^{\rm (r)}) + i 
(\eta - \eps_\alpha^{\rm (i)})  } \right]  \no \\
&=&
\frac{1}{\pi}  \frac{\eta - \eps_\alpha^{\rm (i)}}{ (\eps - \eps_\alpha^{\rm (r)})^2 +  
(\eta - \eps_\alpha^{\rm (i)})^2  }.   
\end{eqnarray}
The above expression indicates that 
a negative peak appears ($\delta_{\eta}(\eps - \eps_\alpha) <0 $), 
when the imaginary part of a pole is larger than the value of $\eta$
($\eta < \eps_\alpha^{\rm (i)}$).
Therefore, we can avoid the negative peaks,
when we use a larger value of $\eta$,
which was confirmed among actual numerical calculations 
with the shifted COCG algorithm.  

Finally, 
we comment on the present discussion 
from the interdisciplinary viewpoints
between physics and mathematics;
From the mathematical view point,
we can say that 
the above two numerical aspects disappear,
when the dimension of the Krylov subspace ($\nu$) increases to be enough large. 
We would like to emphasis that, 
even if the dimension is rather small, 
we can obtain fruitful quantitative discussion 
for several physical quantities,
as discussed above. 
In other words, 
mathematics gives a rigorous way (iterative solver) 
to the exact solution and 
physics gives 
a practical measurement of the convergence criteria.

\appendix

\section{Notation used in Section 2}

In Sec.2,  we use the vector notations of 
\begin{eqnarray}
& & | f \rangle \Leftrightarrow (f_1, f_2 .....f_M)^{\rm T} \\
& & \langle g  | \Leftrightarrow (g_1, g_2 .....g_M).
\end{eqnarray}
Particularly, the unit vector of which non-zero component
is only the $i$-th one is denoted as $| i \rangle$;
\begin{eqnarray}
& & | i \rangle \Leftrightarrow (0, 0,...., 1_i, 0,0)^{\rm T}. 
\end{eqnarray}
Inner products are described as
\begin{eqnarray}
& & \langle g | f \rangle \equiv \sum_i g_i f_i
\end{eqnarray}
are
\begin{eqnarray}
& & \langle g | A | f \rangle \equiv \sum_{ij} g_i A_{ij} f_j
\end{eqnarray}
with a $N \times N$ matrix $A$. 
The notation of $  | f \rangle  \langle g | $ indicates a matrix, 
of which a component is given as 
\begin{eqnarray}
\left( | f \rangle  \langle g |  \right)_{ij} =   f_i g_j.
\end{eqnarray}
These notations are 
used in quantum mechanics and called
\lq bra-ket' notation. 
We should say, however, that the above notations 
are slightly different of the original \lq bra-ket' notations.
For example, the original notation of $ \langle g  | $ is 
$\langle g  | \Leftrightarrow (g_1^\ast, g_2^\ast .....g_M^\ast)$. 
The reason of the difference comes from the fact that 
the standard quantum mechanics is given within 
linear algebra with Hermitian matrices 
but the present formulation is not.

\end{document}